\documentclass[12pt]{article}
\usepackage{amsmath}
\usepackage{amsmath,bm}
\usepackage{url}
\usepackage{tikz}
\usetikzlibrary{shapes.geometric}
\usetikzlibrary{arrows}
\usepackage{amsmath}
\usepackage{amssymb}
\usepackage{hyperref}
\usepackage{rotating}
\usepackage{physics}
\usepackage{epsfig}
\usepackage{graphicx}
\usepackage{color}
\usepackage{cite}
\usepackage[font=small,labelfont=bf]{caption}
\usepackage{xcolor}

\newcommand{\beq}{\begin{equation}}
\newcommand{\eeq}{\end{equation}}
\newcommand{\ga}{\lower.7ex\hbox{$\;\stackrel{\textstyle>}{\sim}\;$}}
\newcommand{\la}{\lower.7ex\hbox{$\;\stackrel{\textstyle<}{\sim}\;$}}

\hypersetup{
    colorlinks = true,
    citecolor = {blue},
    linkcolor = {blue},
    urlcolor = {blue},
}

\begin{document}

\def\jcap{\ref@jnl{J. Cosmology Astropart. Phys.}}

\begin{flushright}
{\tt KCL-PH-TH/2018-69}, {\tt CERN-TH-2018-260}  \\
{\tt ACT-04-18, MI-TH-1813} \\
{\tt UMN-TH-3806/18, FTPI-MINN-18/21} \\
\end{flushright}

\vspace{0.2cm}
\begin{center}
{\bf {\large A General Classification of Starobinsky-like Inflationary  \\[0.2cm] Avatars of SU(2,1)/SU(2)$\times$U(1) No-Scale Supergravity }}

\end{center}
\vspace{0.1cm}

\begin{center}{
{\bf John~Ellis}$^{a}$,
{\bf Dimitri~V.~Nanopoulos}$^{b}$,
{\bf Keith~A.~Olive}$^{c}$ and
{\bf Sarunas~Verner}$^{c}$}
\end{center}

\begin{center}
{\em $^a$Theoretical Particle Physics and Cosmology Group, Department of
  Physics, King's~College~London, London WC2R 2LS, United Kingdom;\\
Theoretical Physics Department, CERN, CH-1211 Geneva 23,
  Switzerland}\\[0.2cm]
{\em $^b$George P. and Cynthia W. Mitchell Institute for Fundamental
 Physics and Astronomy, Texas A\&M University, College Station, TX
 77843, USA;\\ 
 Astroparticle Physics Group, Houston Advanced Research Center (HARC),
 \\ Mitchell Campus, Woodlands, TX 77381, USA;\\ 
Academy of Athens, Division of Natural Sciences,
Athens 10679, Greece}\\[0.2cm]
{\em $^c$William I. Fine Theoretical Physics Institute, School of
 Physics and Astronomy, University of Minnesota, Minneapolis, MN 55455,
 USA}
 
 \end{center}

\vspace{0.1cm}
\centerline{\bf ABSTRACT}
\vspace{0.1cm}

{\small Measurements of the cosmic microwave background (CMB) favour models of inflation
with a small tensor-to-scalar ratio $r$, as predicted by the Starobinsky $R + R^2$ model. It has
been shown previously that various models based on no-scale supergravity with different forms of
superpotential make predictions similar to those of the Starobinsky model. In this paper we present a unified and general treatment
of Starobinsky avatars of no-scale supergravity, using the underlying non-compact SU(2,1)/SU(2)$\times$U(1)
symmetry to demonstrate equivalences between different models, exhibiting 6 specific equivalence classes.
 }

\vspace{0.2in}

\begin{flushleft}
December 2018
\end{flushleft}
\medskip
\noindent

\newpage

\section{Introduction}

Measurements of the cosmic microwave background (CMB) \cite{planck18} continue to be consistent with the
inflationary paradigm \cite{reviews}. So far, the perturbations in the CMB do not exhibit any deviations from 
Gaussianity, and the spectrum of scalar perturbations is almost scale-invariant, with a small tilt:
$n_s \sim 0.96$ to 0.97 \cite{planck18}. To date, there is only an upper limit on the tensor-to-scalar ratio, $r \lesssim 0.06$ \cite{rlimit}.
In combination, these constraints in the $(n_s, r)$ plane exclude many models of inflation
that have been considered in the past, such as monomial single-field models. However, one
simple model is consistent with all the experimental measurements, namely the original
Starobinsky $R + R^2$ model \cite{Staro}.

In view of the virtues of low-energy supersymmetry \cite{Nilles:1983ge,Haber:1984rc}, such as making the hierarchies between the electroweak,
inflationary and gravitational scales less unnatural~\cite{natural}, providing a plausible candidate for cold dark matter~\cite{EHNOS},
facilitating the unification of the strong, weak and electromagnetic interactions~\cite{GUTs}, and predicting successfully the
mass of the Higgs boson~\cite{Hmass} and that its couplings should resemble those in the Standard Model~\cite{Hcouplings}, we think that
an inflationary model should be (approximately) supersymmetric \cite{cries}. In the context of cosmology, one must combine 
supersymmetry with gravity, so the appropriate framework is supergravity \cite{Nilles:1983ge}. Moreover, in order to avoid
unacceptable anti-de Sitter minima in the effective potential, we are led to consider no-scale supergravity \cite{no-scale,Ellis:1983sf}~\footnote{For 
a review of early work on no-scale supergravity, see~\cite{LN}.},
as appears generically in the effective low-energy limit of string compactifications \cite{Witten}. The question then arises
whether one can find in the no-scale framework models that yield Starobinsky-like predictions for the
inflationary observables.

A positive answer to this question was found in~\cite{ENO6}, where it was shown that the simplest 
Wess-Zumino form of superpotential, $W = \mu \phi^2 + \lambda \phi^3$, would yield Starobinsky-like inflation
for suitable values of the superpotential parameters $\mu$ and $\lambda$. An alternative realization of
the $R + R^2$ theory in the framework of no-scale supergravity, with a very different form of superpotential,
had been given previously in~\cite{Cecotti}, though without discussing the connection to models of inflation. 
Following~\cite{ENO6}, a number of other Starobinsky-like inflationary models were obtained as avatars of
no-scale supergravity in~\cite{Avatars}. However, the existence of an underlying connection between all
these no-scale realizations of Starobinsky-like inflation remained an open question.

Here we address this question systematically using the underlying symmetries of no-scale supergravity.
The original minimal no-scale supergravity model contained a single chiral field $T$, with dynamics
described by the K\"ahler potential
\begin{equation}
K \; = \; - \, 3 \ln (T + \bar{T}) \, .
\label{Minimal}
\end{equation}
The field $T$ parameterizes a non-compact SU(1,1)/U(1) coset manifold \cite{no-scale,Ellis:1983sf,EKN1}, and $\bar{T}$
is its complex conjugate. As was shown in~\cite{Witten}, 
this type of K\"ahler structure emerges naturally in simple compactifications of string theory, where $T$ 
can be identified as a modulus field. Generalizations of the minimal no-scale model (\ref{Minimal}) can
be constructed with higher-dimensional non-compact coset manifolds such as 
SU(N,1)/SU(N)$\times$U(1)~\cite{EKN2}~\footnote{There are also generalizations of (\ref{Minimal}) 
in which the supergravity fields parameterize a direct product of such non-compact coset manifolds, 
but we do not discuss them here.}. As was discussed in~\cite{Avatars}, it is not possible to construct a
Starobinsky-like model using the minimal no-scale K\"ahler potential (\ref{Minimal}) {\footnote{However, it is still possible to construct viable inflationary models with a de Sitter plateau in the minimal no-scale supergravity, and such models were considered in~\cite{GL, EENOS,rs}.}}, and the models
constructed in~\cite{Cecotti,ENO6,KLno-scale,FKR,Avatars} were based on a non-compact SU(2,1)/SU(2)$\times$U(1) coset 
manifold.

This theory may be parameterized in terms of two chiral fields either of two equivalent forms:
\begin{equation}
K \; = \; - \, 3 \ln (T + \bar{T} - \frac{|\phi|^2}{3}) \qquad {\rm or} \qquad K \; = \; - \, 3 \ln (1 - \frac{|y_1|^2}{3} - \frac{|y_2|^2}{3}) \, .
\label{SU21}
\end{equation}
One can transform between these two equivalent forms using the underlying non-compact SU(2,1)/SU(2)$\times$U(1)
symmetry. When one does so, the superpotential is in general modified. In this paper we study how
SU(2,1)/SU(2)$\times$U(1) transformations can be used to exhibit equivalences between Starobinsky-like
inflationary models with superpotentials that appear {\it a priori} to be distinct.

The layout of this paper is as follows. In Section~\ref{Framework} we review the general structure of no-scale
supergravity with a view to the subsequent analysis. In Section~\ref{21} we focus on the SU(2,1)/SU(2)$\times$U(1) case
 and identify four branches of Starobinsky-like models, which are specified by fixing one of the fields
parameterizing the non-compact coset manifold and making a suitable canonical field redefinition,
as illustrated in Fig.~\ref{fig:branches}. These branches are then related by simple field transformations,
as illustrated in Fig.~\ref{fig:branches2}. Then, in Section~\ref{SuperStaro} we consider the most general form of third-order superpotential that
reproduces a Starobinsky-like inflationary potential in SU(2, 1)/SU(2)$\times$U(1) no-scale supergravity models. 
Within each branch, Starobinsky-like models can be grouped into six
classes, as shown in Eqs.~(\ref{cases1}) and (\ref{cases2}). This general classification is illustrated in Section~\ref{examples} by some
Starobinsky-like models known previously~\cite{Cecotti,ENO6,Avatars}, as well as some new examples.
Finally, in Section~\ref{conx} we summarize our results and draw some general conclusions.

\section{No-Scale Supergravity Framework}
\label{Framework}

We first recall some of the essential features of supergravity models, which we interpret as low-energy effective theories at energies significantly smaller than the Planck scale. The geometric properties of the scalar field space are characterized by the K\"ahler potential $K(\Phi_i, \bar{\Phi}_{\bar{j}})$, where the fields $\Phi_i$ are complex scalar fields and the fields $\bar{\Phi}_{\bar{j}}$ are their Hermitian conjugates. The kinetic terms of the Lagrangian are given by:
\begin{equation}\label{genkin}
{\cal L}_{kin} = K^{i \bar{j}} \partial_{\mu} \Phi_i \partial^{\mu} \bar{\Phi}_{\bar{j}} \, ,
\end{equation}
where $K^{i \bar{j}} \equiv \partial^2 K/\partial_{\mu}{\Phi}_i \partial^{\mu}{\bar{\Phi}}_{\bar{j}}$ is  the K\"ahler metric. To define the supergravity dynamics of the complex scalar fields $\Phi_i$, one introduces the superpotential $W$, which is a holomorphic function of the $\Phi_i$. The K\"ahler function is then defined as $G \equiv  K + \ln W + \ln \overline{W}  $, and the corresponding supergravity action can be expressed as:
\begin{equation}
S = \int d^4 x \sqrt{-g} \left[K^{i \bar{j}} \partial_{\mu} \Phi_i \partial^{\mu} \bar{\Phi}_{\bar{j}} - V  \right] \, ,
\end{equation}
where the effective scalar potential is
\begin{equation}\label{genpot}
V = e^G \left[ \pdv{G}{\Phi_i} K_{i \bar{j}} \pdv{G}{\bar{\Phi}_{\bar{j}}} - 3 \right],
\end{equation}
and $K_{i \bar{j}}$ is the inverse K\"ahler metric. 

The next step is to apply the general formalism within a specific supergravity model. In this paper we consider no-scale supergravity, 
which was first described in~\cite{no-scale,Ellis:1983sf} in its minimal version. 
We consider here the following generalized K\"ahler potential for $N$ complex scalar fields~\cite{EKN2}
that parametrize a non-compact SU(N,1)/SU(N)$\times$U(1) coset manifold :
\begin{equation}\label{kahgen}
K = -3 \ln \left(T+\bar{T} - \frac{|\phi_i|^2}{3} \right),
\end{equation}
where $T$ is the complex scalar field which can be associated with the volume modulus in a compactified string model,
and $i = 1, 2,\dots, N-1$. The minimal model with no chiral matter fields has $N=1$, 
in which case the K\"ahler potential~(\ref{kahgen}) can be written in terms of a single volume modulus field $T$,
as in (\ref{Minimal}),
which parametrizes the non-compact SU(1,1)/U(1) coset manifold. Using expressions~(\ref{genkin}) and~(\ref{genpot}),
the kinetic terms of the Lagrangian together with the effective scalar potential are given by:
\begin{equation}
{\cal L}_{kin} = \frac{3}{(T + \bar{T})^2} \partial_{\mu}T \partial^{\mu} \bar{T}
\end{equation}
and
\begin{equation}
V = \frac{\hat{V}}{(T+\bar{T})^2}, \quad \text{where} \quad \hat{V} = \frac{1}{3} (T+\bar{T})	|W_T|^2 - (W \overline{W}_{\bar{T}} + \overline{W} W_{T}) \, .
\end{equation}
The general SU(1,1) isometric transformations for the volume modulus $T$ are given by \cite{EKN1,Avatars}
\begin{equation}\label{isometric}
T \rightarrow \frac{\alpha T + i \beta}{i \gamma T + \delta}, 
\qquad \text{where} \quad \alpha, \beta, \gamma, \delta \in \mathbb{R}  \quad \text{and} \quad \alpha \delta + \beta \gamma = 1 \, .
\end{equation}
Not all the general isometric transformations respect the invariance of the K\"ahler potential.
indeed, only imaginary translations $T \rightarrow T + i \beta$ together with inversions $T \rightarrow \left(\frac{\beta}{\gamma}\right) \frac{1}{T}$ leave it invariant up to a K\"ahler transformation. For example, in the case of an inversion we have $K \to K + f(T) + \bar{f}(\bar{T})$ and $W \to e^{-f} W$ with $f = - \ln \sqrt{\gamma/\beta} T$. However, the superpotential does not remain invariant under these transformation laws and, 
therefore, the effective scalar potential also acquires a new form. Nevertheless, the complex scalar fields should be redefined to obtain
canonically-normalized kinetic terms and for $\gamma/\beta = 4$, we recover the initial form of the effective scalar potential in terms of canonical fields.

The symmetry properties of the non-compact SU(N,1)/SU(N)$\times$U(1) coset space can be understood better by adopting a more symmetric representation~\cite{EKN2}, redefining the corresponding complex scalar fields in the following way:
\begin{equation}\label{symflds}
T = \frac{k}{2} \left(\frac{1 - \frac{y_N}{\sqrt{3}}}{1 + \frac{y_N}{\sqrt{3}}} \right), \qquad \text{and}
\quad
\phi^i = \sqrt{k} \left(\frac{y_{i}}{1 + \frac{y_N}{\sqrt{3}}} \right), \quad \text{with} \quad
i = 1, \dots, N - 1 \, ,
\end{equation}
where $k$ is an arbitrary constant.
The field redefinition~(\ref{symflds}) also transforms the effective superpotential into the following form:
\begin{equation}
W( \Phi^i) \rightarrow \widetilde{W}(y_i) = 
\left[\frac{1}{\sqrt{k}} 
\left(1 + \frac{y_N}{\sqrt{3}} \right) \right]^3 W(y_i), \quad \text{with} \quad
i = 1, \dots, N \, .
\end{equation}
Using these symmetric field redefinitions we obtain the following expression for the non-compact SU(N,1)/SU(N)$\times$U(1) K\"ahler potential~(\ref{kahgen}):
\begin{equation}\label{kahsym}
K = -3 \ln \left(1 - \sum_i \frac{|y_i|^2}{3} \right) \, .
\end{equation}
The symmetric form of the K\"ahler potential~(\ref{kahsym}) informs us that the U(1) phase transformation $y^i \rightarrow e^{i \theta} y^i $
is trivial and can be discarded. Therefore, it is sufficient to consider the transformation laws of the non-compact SU(N,1)/SU(N) coset manifold.
With this simplification, we can define an $N \times N$ complex matrix $U$ that parameterizes the SU(N,1)/SU(N) coset space. The complex matrix $U$ must satisfy the following conditions:
\begin{equation}
U^{\dagger} g U = g, \qquad \text{and} \qquad U^{\dagger} U = I,
\end{equation}
where the diagonal matrix $g$ of the SU(N,1) group is given by:
\begin{equation}
g = diag(\overbrace{1, \dots, 1}^N, -1) \, ,
\end{equation}	
and $I$ is the $N\times N$ identity matrix.
Because the complex matrix $U$ has $\det(U) = 1$, it can be shown that $U \in SL(N, \mathbb{C})$,
and can be related to the projective special linear group $PSL(N, \mathbb{C})$ using the following relation:
\begin{equation}
PSL(N, \mathbb{C}) = \frac{SL(N, \mathbb{C})}{Z(SL(N, \mathbb{C}))} \, ,
\end{equation}
where $Z(SL(N, \mathbb{C}))$ is the center of the special linear group $SL(N, \mathbb{C})$. 
Hence the general invariance laws of the K\"ahler potential
are described by projective linear transformations that are elements of the projective linear group $PSL(N, \mathbb{C})$, 
and the projective linear transformations are given by:
\begin{equation}\label{projlin}
[\phi_1, \dots, \phi_N, 1] \rightarrow 
\left[
\frac{(U \Phi)_1}{(U \Phi)_{N+1}}
, \dots, \frac{(U \Phi)_N}{(U \Phi)_{N+1}}, 1
 \right], \quad \text{with} \quad
 \Phi =
\begin{pmatrix}
\phi_1 \\
\vdots \\
\phi_N\\
1
\end{pmatrix} \, ,
\end{equation}
where the $(U \Phi)_i$ are the respective row vectors.

Returning to the SU(1, 1)/U(1) coset space and redefining the complex scalar field $T$ in terms of variable $y_1$ from the equation~(\ref{symflds}), the K\"ahler potential~(\ref{Minimal}) now acquires the symmetric form
\begin{equation}\label{kahy1}
K = -3 \ln \left(1 - \frac{|y_1|^2}{3} \right) \, .
\end{equation}
To show that the result~(\ref{projlin}) is equivalent to the isometric transformations for the volume modulus $T$~(\ref{isometric}), 
we first find the complex matrix $U$ that parametrizes the $SU(1,1)$ group. The complex $2 \times 2$ matrix $U$ can be expressed as
\begin{equation}
U = 
\begin{pmatrix}
a & \lambda \\
\lambda^* & a
\end{pmatrix}, \quad \text{where} \quad a \in \mathbb{R}_{>0}, \quad \lambda \in \mathbb{C}, \quad a^2 - |\lambda|^2 = 1 \, .
\end{equation}
We find from equation~(\ref{projlin}) that the complex field $y_1$ has the following transformation law:
\begin{equation}\label{y1trans}
\frac{y_1}{\sqrt{3}} \rightarrow \frac{ay_1/\sqrt{3} +\lambda}{\lambda^* y_1/\sqrt{3} + a} 
\end{equation}
and now, if we plug equation~(\ref{y1trans}) back into the symmetric K\"ahler potential~(\ref{kahy1}), we see that it remains invariant. Finally, if we use the field redefinitions~(\ref{symflds}) and transform the field $y_1$ back to the volume modulus field $T$ using $y_1/\sqrt{3} \to (1-2T)/(1+2T)$, we recover successfully the isometric transformations~(\ref{isometric}) with $\alpha = 2 a -2 Re \lambda, \beta = -Im \lambda, \gamma = 4 Im \lambda$, and $\delta = 2a + 2 Re \lambda$.

In the single-field theory based on SU(1,1)/U(1), de Sitter solutions are possible for $W = T^3 -1$ \cite{EKN1,rs,eno9,enno},
corresponding to
$W =  (\sqrt{3} y^3 + 7 y^2 + 9 \sqrt{3} y + 7)/8$ in the symmetric basis. However, this would not lead to a Starobinsky-like 
 inflationary potential~\cite{Avatars}. 
Nevertheless, the mathematical framework introduced in this Section can be applied to non-minimal no-scale supergravity models,
and in next Section we explore how the Starobinsky model of inflation arises in SU(2,1)/SU(2) $\times$ U(1) supergravity models.

\section{SU(2,1)/SU(2)$\times$U(1) No-Scale Supergravity}
\label{21}

We consider now the simplest non-minimal no-scale supergravity model, based on the non-compact 
SU(2,1)/SU(2)$\times$U(1) coset. In this case the K\"ahler 
potential may be written in the following form:
\begin{equation}\label{kah2}
K = -3 \ln \left(T +\bar{T} - \frac{\phi \bar{\phi}}{3} \right) \, ,
\end{equation}
where now the complex scalar fields $(T, \phi)$ parametrize the non-compact SU(2,1)/SU(2) $\times$ U(1) coset manifold. Using the K\"ahler potential~(\ref{kah2}) in conjunction with expression~(\ref{genkin}), we find that the kinetic terms of the Lagrangian are given by:
\begin{equation}\label{kin2}
{\cal L}_{kin} = 
\begin{pmatrix}
\partial_{\mu} \bar{\phi},~\partial_{\mu} \bar{T}
\end{pmatrix}
\left(\frac{3}{
(T+\bar{T}-|\phi|^2/3)^2} \right)
\begin{pmatrix}
\frac{T + \bar{T}}{3} & - \phi \\
-\bar{\phi} & 1
\end{pmatrix}
\begin{pmatrix}
\partial^{\mu} \phi \\
\partial^{\mu} T
\end{pmatrix} \, .
\end{equation}
The effective potential~(\ref{genpot}), which in this case is expressed in terms of a general superpotential $W(T, \phi)$, becomes:
\begin{equation}\label{pot2}
V = \frac{\hat{V}}{\left(T + \bar{T} - \frac{\phi \bar{\phi}}{3}\right)^2} \, ,
\end{equation}
with
\begin{equation}
\hat{V} \equiv \left| \frac{\partial W}{\partial \phi} \right|^2 + \frac{1}{3}(T + \bar{T}) |W_T|^2 + \frac{1}{3} \left(W_T ( \overline{W}_{\bar{\phi}} \bar{\phi} - 3 \overline{W}) + h.c. \right) \, ,
\end{equation}
where $W_T = \partial{W}/ \partial{T}$ and $W_{\phi} = \partial{W}/ \partial{\phi}$. 

Because the general SU(2,1)/SU(2)$\times$U(1) no-scale models are parametrized by two complex scalar fields $(T, \phi)$, or by four real scalar fields, in order to recover the Starobinsky inflationary potential it will be necessary to fix one of the complex scalar fields and hence break the coset space symmetry. Thus, our first step toward deriving the expressions for general 
superpotentials $W(T, \phi)$ capable of yielding a Starobinsky inflationary potential is to find all possible canonically-normalized scalar field expressions. For convenience, we expand fully the kinetic terms~(\ref{kin2}) and fix either field $T$ or $\phi$, in which case
the kinetic cross-terms can be discarded, leaving us with
\begin{equation}\label{kindyn}
{\cal L}_{kin} = \frac{3}{\left(T + \bar{T} - \frac{\phi \bar{\phi}}{3} \right)^2} \partial_{\mu}T \partial^{\mu} \bar{T}
+
\frac{(T+\bar{T})}{\left(T + \bar{T} - \frac{\phi \bar{\phi}}{3} \right)^2} \partial_{\mu}\phi \partial^{\mu} \bar{\phi} \, .
\end{equation}

Let us first follow the same treatment as in~\cite{ENO6} and assume that the $T$ field is fixed, 
with a vacuum expectation value of $\langle Re~T \rangle = \frac{k}{2}$ and $\langle Im~T \rangle = 0$, in which case the kinetic terms~(\ref{kindyn}) become:	
\begin{equation}\label{kinphi}
{\cal L}_{kin} = \frac{k}{\left(k - \frac{\phi \bar{\phi}}{3} \right)^2} \partial_{\mu}\phi \partial^{\mu} \bar{\phi} \, .
\end{equation}
In order to understand better how to normalize canonically the field $\phi$, we assume that the 
imaginary part of the matter field $\phi$ is fixed to $\langle Im~\phi \rangle = 0$ by the dynamics of the potential. The details of this assumption will be discussed later.

Then we find from equation~(\ref{kinphi}) the following redefinition for a canonically-normalized field:
\begin{equation}\label{phigen}
\phi = \pm \sqrt{3 k} \tanh(\frac{x}{\sqrt{6}}) \, ,
\end{equation}
where $x$ is a real scalar field. When considering the general expression for the superpotential  $W(T, \phi)$
that is to yield a Starobinsky inflationary potential, it is crucial to respect the canonical formulation of the model
by using the redefinition~(\ref{phigen}) of the matter field $\phi$.

\begin{figure}[!h]
\begin{center}
\resizebox {13cm} {!} {
\begin{tikzpicture}[ ->,>=stealth',shorten >=1pt,auto,node distance=5.4cm,
                    thick,main node/.style={align=center,circle,draw,font=\linespread{0.8}\sffamily\Large\bfseries}],	
                    \tikzset{
    pil/.style={
           ->,
           line width=0.5mm,
           shorten <=2pt,
           shorten >=2pt,}
},    			 
  \node[main node, rectangle] [fill=yellow!50, yshift=-25mm] (1) {\normalsize Canonical \\ \normalsize Field Redefinition};
  \node[diamond, draw] (I)[fill=green!50, left of=1, xshift=-18mm]{\large \textbf{Branch I~} };
  \node[diamond, draw] (II)[fill=green!50, right of=1,  xshift=18mm]{\large \textbf{Branch II}};
  \node[main node] (2) [fill=red!50, below of=1, yshift=15mm] {SU(2, 1)/ \\ SU(2)$\times$U(1)};
  \node[main node, rectangle] (3) [fill=yellow!50, below of=2, yshift=15mm] {\normalsize Canonical \\ \normalsize Field Redefinition};
    \node[diamond, draw] (III)[fill=green!50, left of=3, xshift=-18mm]{\large \textbf{Branch III} };
  \node[diamond, draw] (IV)[fill=green!50, right of=3, xshift=18mm]{\large\textbf{Branch IV} };

  \path[every node/.style={font=\sffamily\small}]
    (1) edge [pil, right, above]  node {\footnotesize $~~\phi = - \sqrt{3k} \tanh(\frac{x}{\sqrt{6}})$}  (II)  
    edge [pil, left, above] node {\footnotesize $\phi = + \sqrt{3k} \tanh(\frac{x}{\sqrt{6}})$}  (I)
    (2) edge [pil, right] node  {Fixing the field $\langle T \rangle = \frac{k}{2}$}
    (1) edge [pil, right] node {Fixing the field $\langle \phi \rangle=0$} (3)
    
    (3) edge [pil, right, above] node {\footnotesize $T = \frac{k}{2} e^{-\sqrt{\frac{2}{3}} \rho}$}  (IV)  
    edge [pil, left, above] node  {\footnotesize $T = \frac{k}{2} e^{+\sqrt{\frac{2}{3}} \rho}$}  (III)
    (I) edge [pil,-, bend right=-40] (II)
    (I) edge [pil,-, bend left=-30] (III)
    (III) edge [pil,-, bend right=40] (IV)
    (II) edge [pil,-, bend left=30] (IV)
    ;
\end{tikzpicture}}
\end{center}
\caption{\it This diagram shows how, starting from a general two-field superpotential $W(T, \phi)$,
it is possible to break the SU(2,1)/SU(2)$\times$U(1) symmetry by fixing one of the complex scalar fields 
and then casting the dynamical field in canonical form by a suitable field redefinition, yielding four
distinct branches of models with Starobinsky-like effective scalar potentials.} \label{fig:branches}
\end{figure}
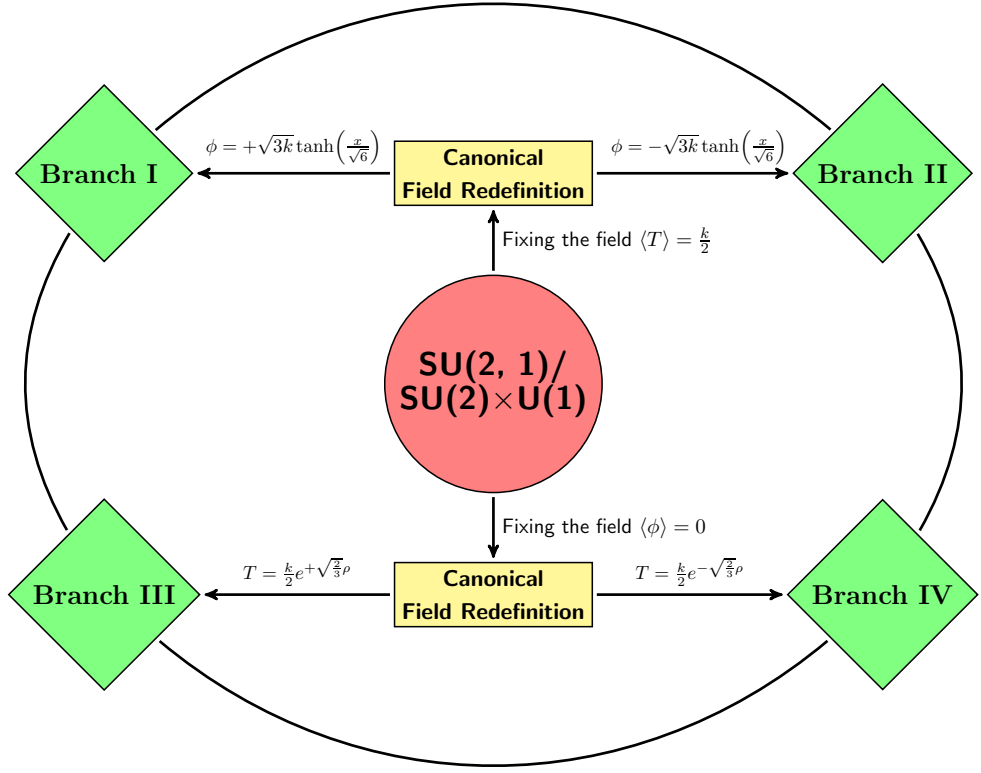

Alternatively, instead of fixing the field $T$, one can perform the analogous procedure of setting the vacuum expectation value of 
the matter field to $\langle \phi \rangle = 0$. In this case the kinetic terms of the Lagrangian~(\ref{kindyn}) are expressed as:
\begin{equation}\label{kinT}
{\cal L}_{kin} = \frac{3}{\left(T + \bar{T} \right)^2} \partial_{\mu}T \partial^{\mu} \bar{T} \, .
\end{equation}
Similarly, we assume a vacuum expectation value for $T$ with $\langle Im~T \rangle = 0$, 
so that the volume modulus $T$ is now a real field and the redefined canonically-normalized field can be expressed as:
\begin{equation}\label{Tgen}
T = \frac{k}{2} e^{\pm \sqrt{\frac{2}{3}} \rho} \, ,
\end{equation}
where the field $\rho$ is now real and the coefficient in front of~(\ref{kinT}) is chosen to be compatible with the symmetric field redefinitions~(\ref{symflds}). Hence we can see that, by fixing one of the complex scalar fields $(T, \phi)$ and then performing 
a canonical field redefinition~(\ref{kinphi}) or~(\ref{kinT}), the SU(2,1)/SU(2)$\times$U(1) symmetry can be broken into 
one of four different branches according to the breaking diagram shown in Fig.~\ref{fig:branches}.

When seeking models leading to the Starobinsky inflationary potential, one may consider any of the four possible branches shown in Fig.~\ref{fig:branches}. 
However, it is possible to perform a SU(2,1)/SU(2)$\times$U(1) projective linear transformation on a general superpotential with fully dynamical fields to 
connect the different superpotential branches, as we now show following the description introduced previously.
To show this, we adopt the symmetric approach and switch from the basis $(T, \phi)$ to the basis $(y_1, y_2)$ using the general field redefinitions~(\ref{symflds}),
the following relations:
\begin{equation}\label{symfld1}
y_1 =\sqrt{k} \left(\frac{2 \phi}{k + 2T}\right), \qquad y_2 = \sqrt{3} \left( \frac{k - 2T}{k + 2T} \right) \, .
\end{equation}
and the inverse relations:
\begin{equation}\label{symfld2}
T = \frac{k}{2} \left(\frac{1 - y_2/\sqrt{3}}{1 + y_2/\sqrt{3}} \right), \qquad \phi =\sqrt{k} \left( \frac{y_1}{1 + y_2/\sqrt{3}} \right) \, .
\end{equation}
In the $(y_1, y_2)$ basis the K\"ahler potential~(\ref{kahgen}) takes the symmetric form
\begin{equation}\label{kah3}
K = -3 \ln \left(1 - \frac{|y_1|^2 + |y_2|^2}{3} \right) \, .
\end{equation}
The expressions for the canonically field redefinitions~(\ref{phigen}) and~(\ref{Tgen}) in the $(y_1, y_2)$ symmetric basis are the following in
the different branches:
\begin{small}
\begin{align}\label{symtrans1}
& \textbf{Branch I:}~\left(\phi = +\sqrt{3k} \tanh(\frac{x}{\sqrt{6}});~\langle T \rangle = \frac{k}{2} \right) \longrightarrow
\left(y_1 = +\sqrt{3} \tanh(\frac{x}{\sqrt{6}});~\langle y_2 \rangle = 0 \right) \, , \\ \label{symtrans2}
& \textbf{Branch II:}~\left(\phi = -\sqrt{3k} \tanh(\frac{x}{\sqrt{6}});~\langle T \rangle = \frac{k}{2} \right) \longrightarrow
\left(y_1 =- \sqrt{3} \tanh(\frac{x}{\sqrt{6}});~\langle y_2 \rangle = 0 \right) , \\ \label{symtrans3}
& \textbf{Branch III:}~\left( \langle \phi \rangle = 0;~ T = \frac{k}{2} e^{+ \sqrt{\frac{2}{3}} \rho} \right) \longrightarrow
\left(\langle y_1  \rangle = 0;~ y_2  = -\sqrt{3} \tanh(\frac{\rho}{\sqrt{6}}) \right) \, , \\ \label{symtrans4}
& \textbf{Branch IV:}~\left( \langle \phi \rangle = 0;~ T = \frac{k}{2} e^{- \sqrt{\frac{2}{3}} \rho} \right) \longrightarrow
\left(\langle y_1  \rangle = 0;~ y_2  = +\sqrt{3} \tanh(\frac{\rho}{\sqrt{6}}) \right) \, .
\end{align}
\end{small}
After recovering the canonically-normalized kinetic terms in the $(y_1, y_2)$ symmetric basis using
equations~(\ref{symtrans1})-(\ref{symtrans4}), we consider general superpotential expressions $W(y_1, y_2)$.

\begin{figure}[!h]
\begin{center}
\resizebox{\textwidth}{!}{
\begin{tikzpicture}[->,>=stealth',shorten >=1pt,auto,node distance=6.5cm,
                    thick,main node/.style={align=center,circle,draw,font=\linespread{0.8}\sffamily\Large\bfseries},Arrow/.style = {line width=2mm, draw=gray, 
                -{Triangle[length=3mm,width=4mm]},
                shorten >=1mm, shorten <=1mm}],
                    	
\tikzset{
    pil/.style={
           ->,
           line width=0.5mm,
           shorten <=2pt,
           shorten >=2pt,}
},               
  \node[fill=red!50, main node] (1) {Effective \\ Scalar \\ Potential};			
  \node[fill=green!50, main node, ellipse, draw] (I)[above left of=1]{Branch I \\ \scriptsize Superpotential $\mathbf{W_I(y_1, y_2)}$};
  \node[fill=green!50, main node, ellipse, draw] (II)[above right of=1] {Branch II \\ \scriptsize Superpotential $\mathbf{W_{II}(y_1, y_2)}$};
 
    \node[main node, ellipse, draw] (III)[fill=green!50, below left of=1] {Branch III \\ \scriptsize Superpotential $\mathbf{W_{III}(y_1, y_2)}$};
  \node[fill=green!50, main node, ellipse, draw] (IV)[below right of=1]{Branch IV \\ \scriptsize Superpotential $\mathbf{W_{IV}(y_1, y_2)}$};
  \path[every node/.style={font=\sffamily\small}]
    (I) edge [pil, right, above] node[above, rotate=-45] {\scriptsize $\mathbf{~y_1 = +\sqrt{3} \tanh(\frac{x}{\sqrt{6}})}$} node[below, rotate=-45] {\scriptsize $\mathbf{\langle y_2 \rangle = 0}$}(1)
         edge [pil,<->, bend left=30] node[above] {Field Transformation} node[below] {$\mathbf{y_1 \leftrightarrow -y_1}$}node[below, yshift=-5mm] {$\mathbf{y_2 \leftrightarrow -y_2}$}node[below, yshift=-10mm] {$\bm{\alpha=-1};\bm{\beta = 0}$}(II)

edge [pil,<->, bend right=30] node[above, xshift=-15mm] {\hspace {-0.7cm} Field Transformation} node[below,xshift=-15mm] {$\mathbf{y_1 \leftrightarrow -y_2}$}node[below, yshift=-5mm, xshift=-15mm] {$\mathbf{y_2 \leftrightarrow y_1}$} node[below, yshift=-10mm, xshift=-15mm] {$\bm{\alpha = 0};\bm{\beta = -1}$}(III)         
         
   (II) edge [pil, left, above]  node[above, rotate=45] {\scriptsize $\mathbf{y_1 = -\sqrt{3} \tanh(\frac{x}{\sqrt{6}})}$} node[below, rotate=45] {\scriptsize $\mathbf{\langle y_2 \rangle = 0}$}(1)
   
edge [pil,<->, bend left=30] node[above, xshift=15mm] {\hspace{0.5cm} Field Transformation} node[below,xshift=15mm] {$\mathbf{y_1 \leftrightarrow -y_2}$} node[below, yshift=-5mm, xshift=15mm] {$\mathbf{y_2 \leftrightarrow y_1}$}node[below, yshift=-10mm, xshift=15mm] {$\bm{\alpha = 0};\bm{\beta = -1}$}(IV)     
   
   (III) edge [pil, left, below]  node[above, rotate=45] {\scriptsize $\mathbf{\langle y_1 \rangle = 0}$} node[below, rotate=45] {\scriptsize $\mathbf{~y_2 = -\sqrt{3} \tanh(\frac{\rho}{\sqrt{6}})}$}(1)  
   
edge [pil,<->, bend right=30] node[above, yshift=3mm] {Field Transformation} node[below] {$\mathbf{y_1 \leftrightarrow -y_1}$}node[below, yshift=-5mm] {$\mathbf{y_2 \leftrightarrow -y_2}$} node[below, yshift=-10mm] {$\bm{\alpha=-1};\bm{\beta = 0}$}(IV)
   
   (IV) edge [pil, right, below] node[above, rotate=-45] {\scriptsize $\mathbf{\langle y_1 \rangle = 0}$} node[below, rotate=-45] {\scriptsize $\mathbf{y_2 = +\sqrt{3} \tanh(\frac{\rho}{\sqrt{6}})}$}(1);
\end{tikzpicture}}
\end{center}
\caption{\it This diagram shows the transformation laws between the superpotentials in different branches, together with the field fixings and canonical field redefinitions that yield the same Starobinsky-like effective scalar potential.} \label{fig:branches2}
\end{figure}
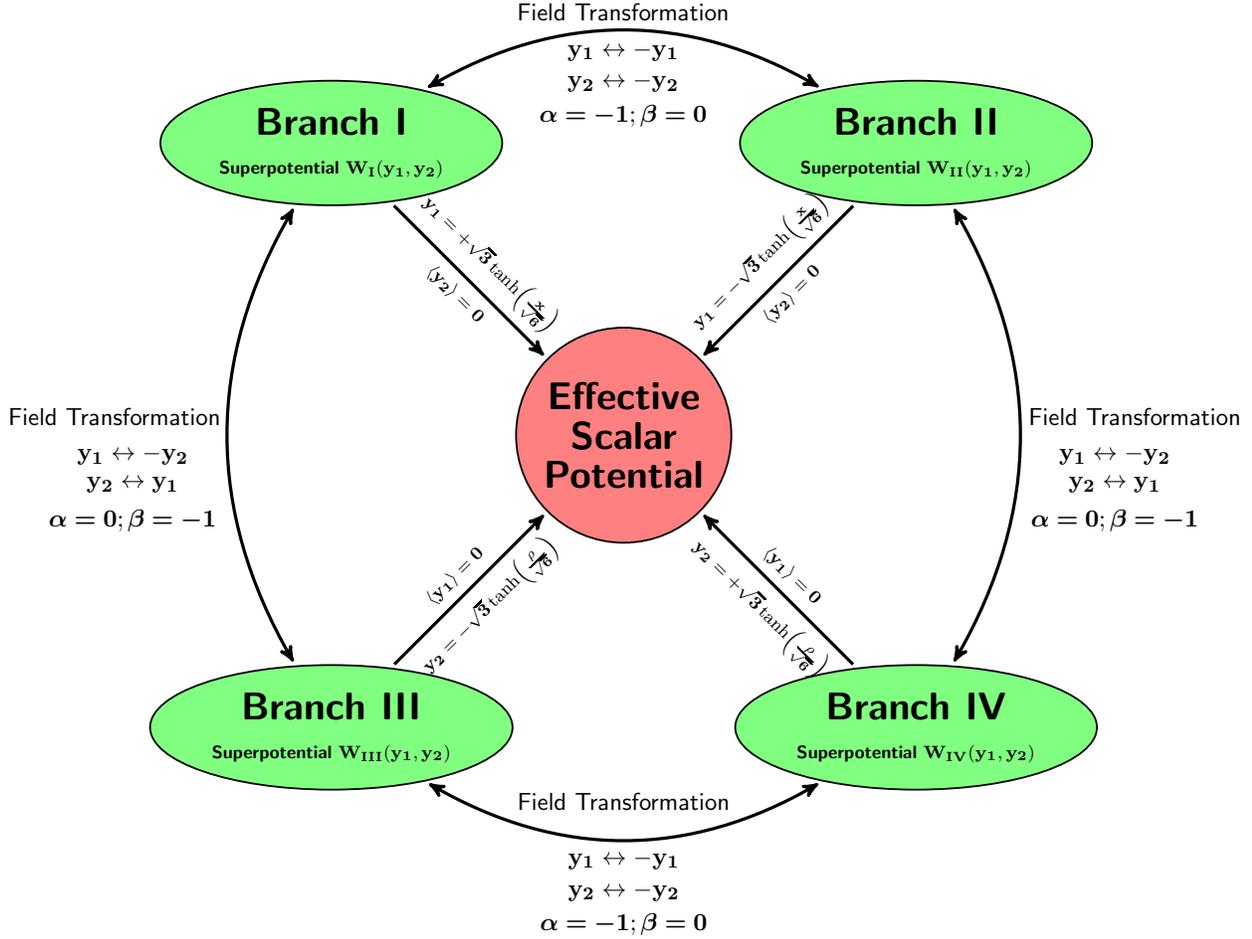

The SU(2,1)/SU(2)$\times$U(1) coset space can be parameterized with the following complex matrix $U$:
\begin{equation}\label{compmat}
U =
\begin{pmatrix}
\alpha & \beta & 0 \\
-\beta^* & \alpha^* & 0 \\
0 & 0 & 1
\end{pmatrix}, \quad \text{where} \quad \alpha, \beta \in \mathbb{C}, \quad |\alpha|^2 + |\beta|^2 = 1 \, .
\end{equation}
Using $U$~(\ref{compmat}) together with equation~(\ref{projlin}), we obtain the following transformation laws for the fields $y_1$ and $y_2$:
\begin{equation}\label{syminv}
y_1 \rightarrow \alpha y_1 + \beta y_2, \qquad y_2 \rightarrow -\beta^* y_1 + \alpha^* y_2 \, .
\end{equation}
When we apply the transformation laws~(\ref{syminv}), the K\"ahler potential~(\ref{kah3}) remains invariant while the general superpotential $W(y_1, y_2)$ transforms non-trivially. It is usually more convenient to work in the symmetric $(y_1, y_2)$ basis when considering the general superpotential at the starting-point of the analysis. The essence of our approach is to start with a general superpotential $W(y_1, y_2)$, corresponding to one of the four different branches in Fig.~\ref{fig:branches}, and then apply the transformation laws~(\ref{syminv}) to obtain the corresponding superpotential in  a different branch. It is crucial to note that all the transformations must be performed for dynamical fields, following which three of the four real fields are fixed while the remaining field is put into the canonical form according to the branch rules. The transformation relations between different branch superpotentials in the $(y_1, y_2)$ symmetric basis are illustrated in Fig~\ref{fig:branches2}.

If one specific superpotential form is known, by performing the indicated field transformations one can find the corresponding
superpotential in a different branch. Thus, for SU(2,1)/SU(2)$\times$U(1) no-scale supergravity models there are four different forms of superpotential $W(y_1, y_2)$ that, by following the corresponding branch parametrization rules~(\ref{symtrans1})-(\ref{symtrans4}), yield identical effective scalar potentials. In the next Section we find all four different branch superpotential expressions that reduce to the Starobinsky inflationary potential in one of the two real fields. The same procedure could also be followed to recover a different effective scalar potential, and analogous transformation rules for general SU(N,1)/SU(2)$\times$U(1) no-scale supergravity models could also be derived.

\section{Starobinsky Superpotentials: General Classification}
\label{SuperStaro}


We consider now the most general superpotential that allows us to recover the Starobinsky inflationary potential in SU(2,1)/SU(2)$\times$U(1) no-scale supergravity models. As was mentioned in the previous Section, by applying the transformation laws~(\ref{symtrans1})-(\ref{symtrans4}), 
which are also depicted in Fig~\ref{fig:branches2}, one can transform between the different superpotential branches. 

For a general superpotential $W(y_1, y_2)$, we have the following expression for the effective potential:
\begin{equation}\label{effsym}
V = \frac{\hat{V}}{\left(1 - \frac{|y_1|^2 + |y_2|^2}{3} \right)^2} \, ,
\end{equation}
where
\begin{equation}\label{effsym2}
\hat{V} = |W_1|^2 + |W_2|^2 - \frac{1}{3} |3W - W_1 y_1 - W_2 y_2|^2 \, ,
\end{equation}
with $W_1 = \partial W/ \partial y_1$ and $W_2 = \partial W/ \partial y_2$.\\

\noindent
$\bullet$
In order to find the general expressions for all four different superpotential branches that will allow us to recover the Starobinsky inflationary potential with canonically-normalized kinetic terms, we start with a general superpotential expression for Branch I, which can be expressed as:
\begin{align}\label{genstaro1}
& \textbf{Branch I:}~&W(y_1, y_2)& &=& &a y_1+b y_1^2+c y_1^3+d y_2+e y_2 y_1+f y_2 y_1^2 + g(y_1, y_2) \, , 
\end{align}
where the additional term $g(y_1, y_2)$ obeys the following conditions: $g(y_1, 0)
= 0$, $\partial g/ \partial y_1 (y_1 , 0)$  $= 0$ and $\partial g/ \partial y_2 (y_1, 0) = 0$.
Terms containing factors $y_2^n$, with $n> 1$ may also appear in $g$ but, since we will require $\langle y_2 \rangle = 0$, these terms do
not contribute to $V$. We have not included a constant term in (\ref{genstaro1}) (or in the general form for
$W$ for the other branches) to avoid supersymmetry breaking of order the inflationary scale.  \\
~\\
$\bullet$ If we perform the transformation (\ref{syminv}), with $\alpha = -1$ and $\beta = 0$,  we obtain
\begin{align}\label{genstaro2}
&\textbf{Branch II:}~&W(y_1, y_2)& &=& &-a y_1+b y_1^2-c y_1^3-d y_2+e y_2 y_1-f y_2 y_1^2+ g(y_1, y_2) \, .
\end{align} \\
$\bullet$
If instead we apply the transformation with $\alpha = 0$ and $\beta = -1$ to the general expression for Branch I (\ref{genstaro1})
we obtain
\begin{align}\label{genstaro3}
& \textbf{Branch III:}~&W(y_1, y_2)& &=& &-a y_2+b y_2^2-c y_2^3+d y_1-e y_1 y_2+f y_1 y_2^2+ h(y_1, y_2) \, ,
\end{align}
where, as before, the additional term $h(y_1, y_2)$ must now satisfy the following conditions: $h (0, y_2) = 0$, $\partial h/ \partial y_1 (0 ,y_2) = 0$ and $\partial h/ \partial y_2 (0 ,y_2) = 0$, when $\langle y_1 \rangle = 0$.  \\
~\\
$\bullet$
Finally, applying either the same transformation to Branch II or applying the previous transformation  with $\alpha = -1$ and $\beta = 0$  to Branch III,
we obtain
\begin{align}\label{genstaro4}
&\textbf{Branch IV:}~&W(y_1, y_2)& &=& &a y_2+b y_2^2+c y_2^3-d y_1-e y_1 y_2-f y_1 y_2^2+ h(y_1, y_2)\, .
\end{align}

Using the form of Branch I superpotential (\ref{genstaro1}), we can derive $\hat{V}$ from (\ref{effsym2})
and match that with a known solution from \cite{Avatars} in which 
\beq
\hat{V} = |y_1|^2 |1-y_1/\sqrt{3}|^2 \, ,
\eeq
corresponding to the Wess-Zumino model found in \cite{ENO6}. Matching the coefficients leads to the following four sets of solutions:
\begin{small}
\begin{equation}
\begin{cases}\label{cases1}
a = 0, \qquad c = +\frac{b \left(\sqrt{1-4 b^2}-2\right)}{3 \sqrt{3}}, \qquad d = 0, \qquad e = \pm \sqrt{1-4 b^2}, \qquad f =\mp \frac{\sqrt{1-4 b^2}+2 b^2}{\sqrt{3}} \, , \\
a = 0, \qquad c = -\frac{b \left(\sqrt{1-4 b^2}+2\right)}{3 \sqrt{3}}, \qquad d = 0, \qquad e = \pm \sqrt{1-4 b^2},\qquad  f = \mp \frac{\sqrt{1-4 b^2} - 2 b^2}{\sqrt{3}} \, ,
\end{cases}
\end{equation}
\end{small}
\noindent
where all the coefficients are expressed in terms of an arbitrary free parameter $b$. 
There are in addition the two solutions:
\begin{small}
\begin{equation}
\label{cases2}
\hspace{-0.3cm}
b = -\frac{\sqrt{3} a}{2 a^2+3}, \; c = \frac{16 a^6+72 a^4+108 a^2+27}{36 a \left(2 a^2+3\right)^2}, \; d = \pm i a, \;  e = \mp \frac{2 i \sqrt{3} a}{2 a^2+3}, \; f = \mp \frac{i \left(4 a^2 \left(2 a^2+3\right)^2+27\right)}{12 a \left(2 a^2+3\right)^2},
\end{equation}
\end{small}
where now the coefficients are expressed in terms of an arbitrary free parameter $a$.
Eqs.~(\ref{cases1}) and~(\ref{cases2}) encompass all the Branch I solutions that yield the Starobinsky inflationary potential with canonically-normalized kinetic terms. 

\begin{figure}[h!]
\centering
\includegraphics[scale=0.5]{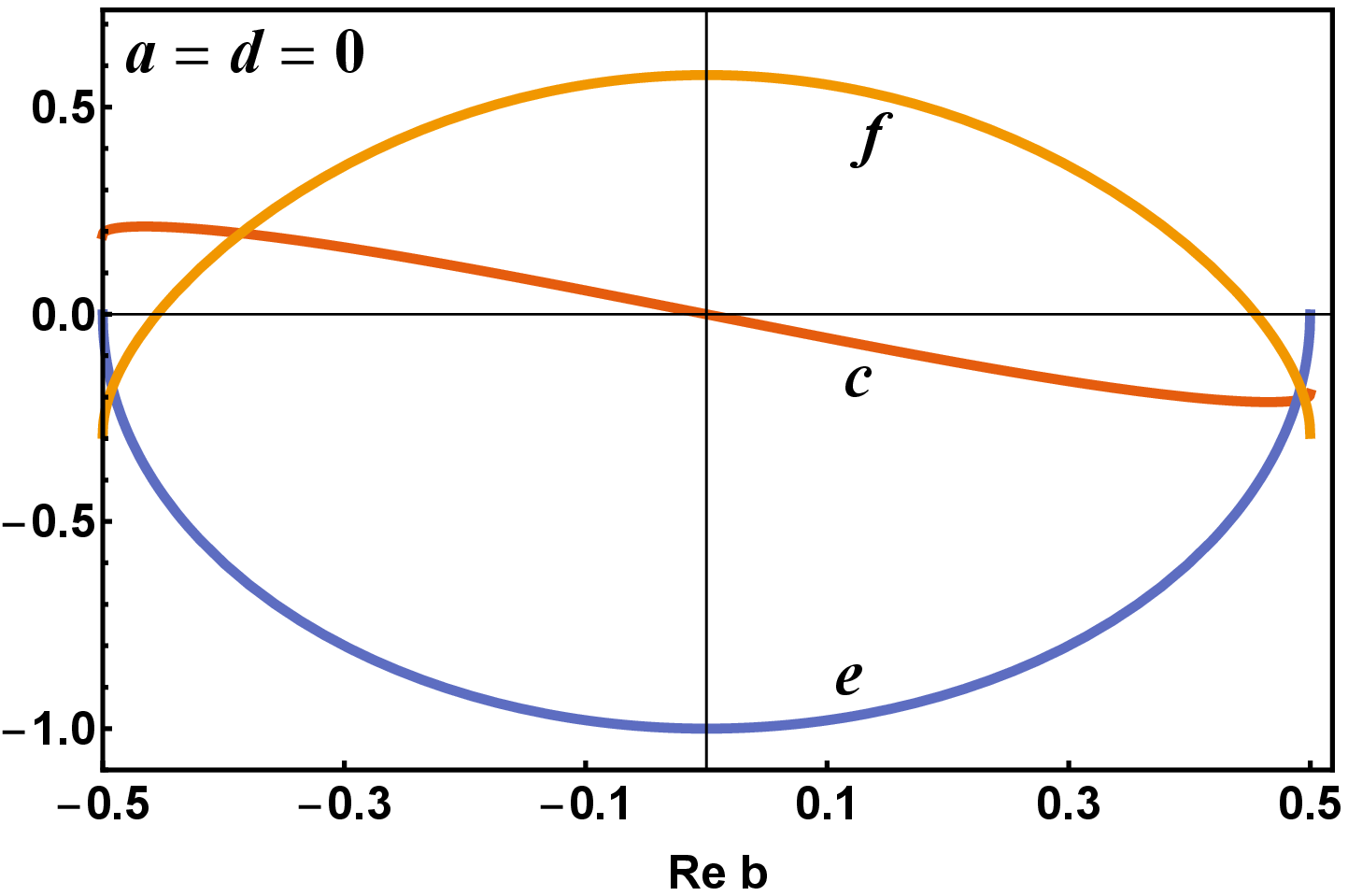}
\hskip .25in
\includegraphics[scale=0.5]{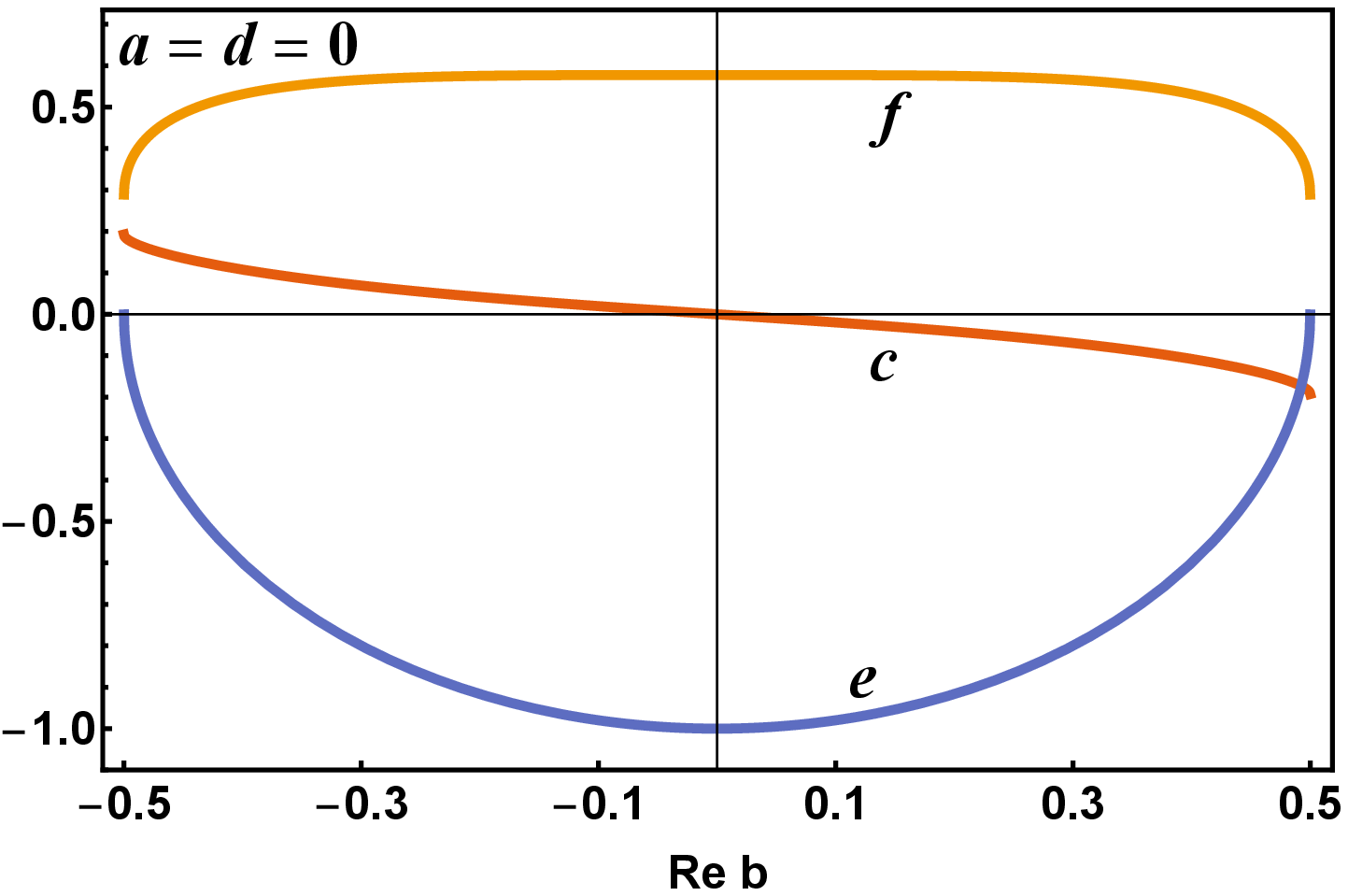}
\caption{\it The superpotential couplings, $c, e$, and $f$ as functions of $b$ that correspond to the Starobinsky solutions
in (\ref{cases1}).  }
\label{real}
\end{figure}

\begin{figure}[h!]
\centering
\includegraphics[scale=0.5]{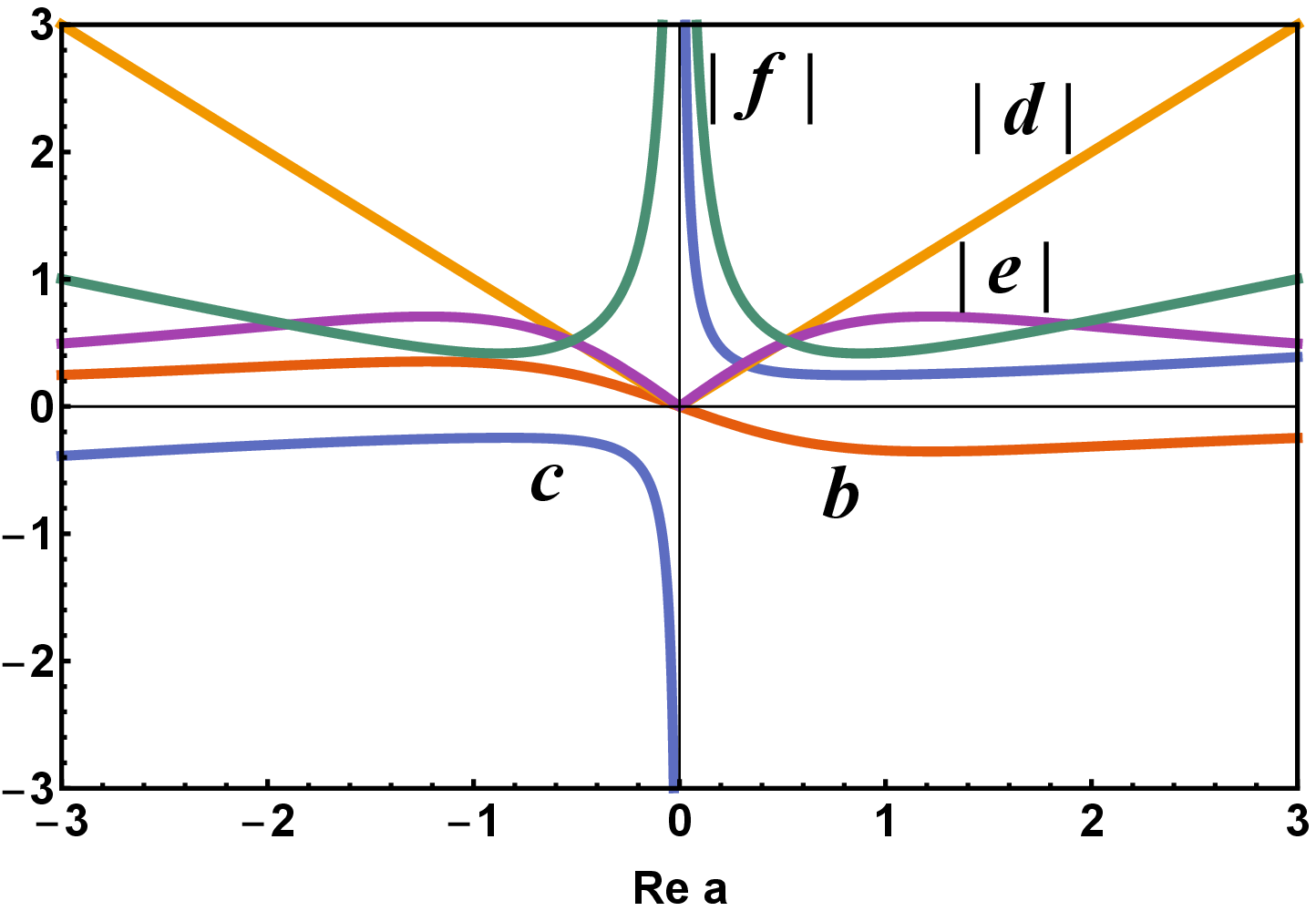}
\hskip .25in
\includegraphics[scale=0.5]{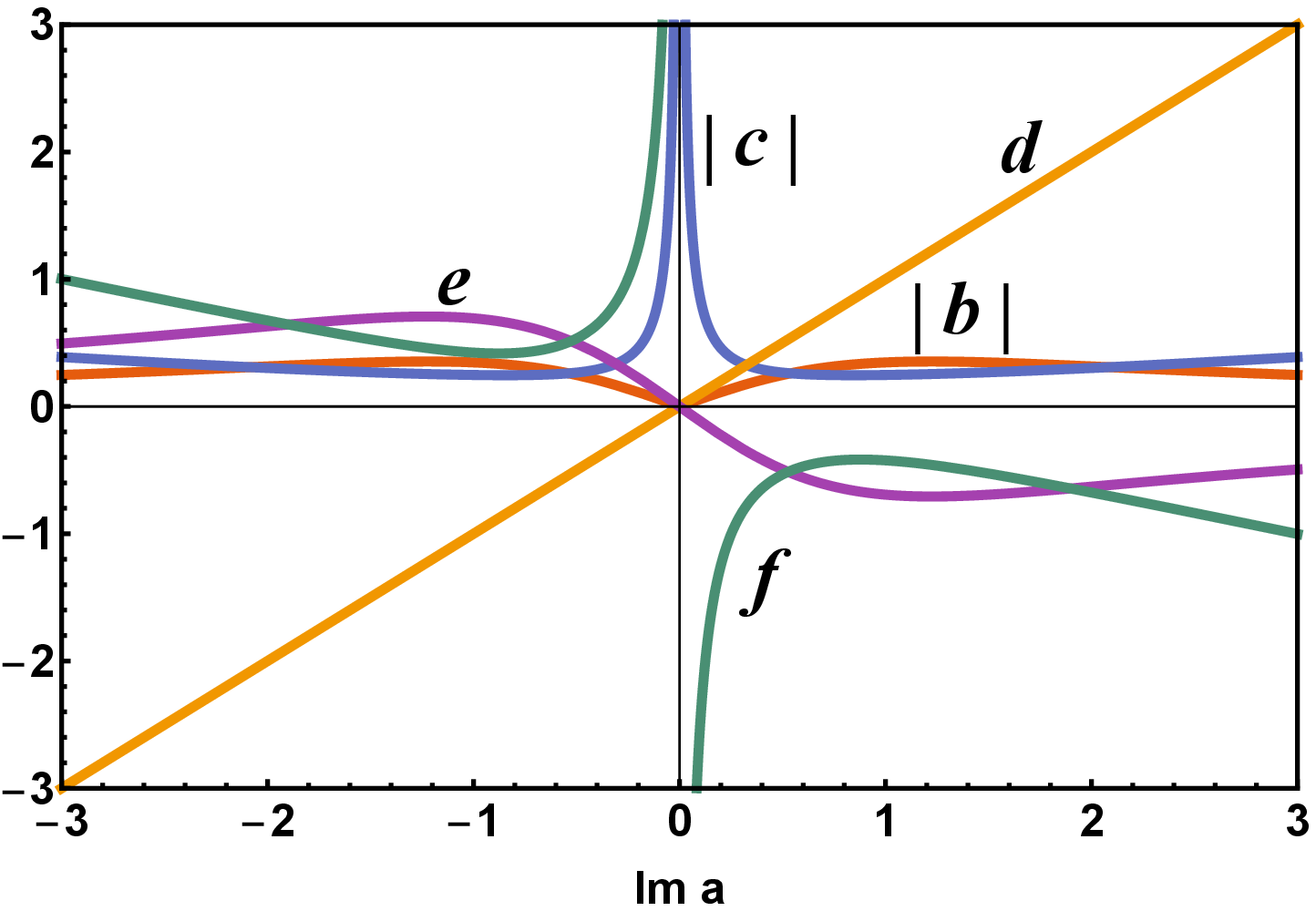}
\caption{\it The superpotential couplings, $b, c, d, e$, and $f$ as functions of $a$ that correspond to the Starobinsky solutions
in (\ref{cases2}).  }
\label{comp}
\end{figure}

For illustration, Figs.~\ref{real} and~\ref{comp} show the couplings necessary for Starobinsky solutions as function of a single free coupling.
In the two panels of Fig. \ref{real}, the free coupling is $b$, corresponding to the solutions in Eq. (\ref{cases1}). We show only two of the solutions,
as the remaining involve only a change in sign for each coupling. In Fig.~\ref{comp}, the free coupling is $a$, corresponding to Eq. (\ref{cases2}).

One can then use the SU(2,1) field transformations given in Eq. (\ref{syminv}) to find the corresponding
Starobinsky-like model in the other branches, as illustrated in Fig.~\ref{fig:branches2}.
For example, if we transform to Branch II, we can reapply the process using now (\ref{genstaro2}) in (\ref{effsym2}) and match to 
\beq
\hat{V} = |y_1|^2 |1+y_1/\sqrt{3}|^2 \, .
\eeq
Solving for the superpotential coefficients $a, b, c, \cdots$, leads to exactly the same sets of solutions as given in Eqs.~(\ref{cases1}) and~(\ref{cases2}).
Similarly using the general superpotentials for Branches III and IV, matching to $\hat{V}$ (with $y_1 \to y_2$),
we again would obtain the solutions in (\ref{cases1}) and~(\ref{cases2}).

Thus, Eqns.~(\ref{genstaro1})-(\ref{genstaro4}), together with (\ref{cases1}) and~(\ref{cases2}) provide all of the solutions
that yield the Starobinsky inflationary potential with canonically-normalized kinetic terms. The solutions (\ref{cases1}) and~(\ref{cases2}) 
provide all solutions within a branch, and the SU(2,1) transformations allow us to rotate between branches with the same form of solutions. 

When considering the inflationary dynamics, we have assumed that three out of four complex scalar fields have been stabilized. To achieve this result, we can include a sufficient stabilization mechanism, which was described in~\cite{Avatars,EKN3}, and tackle the stability problem by adding a quartic stabilization term in the K\"ahler potential that will leave the effective scalar potential unaffected. For Branch I and II solutions we assumed that $\langle y_2 \rangle = 0$, therefore, to stabilize the inflaton field $y_1$, we could add a quartic stabilization term to the K\"ahler potential:
\begin{equation}
K \; = \; - \, 3 \ln (1 - \frac{|y_1|^2}{3} - \frac{|y_2|^2}{3} + \frac{|y_2|^4}{	\Lambda^2}) \, ,
\end{equation}
where we assume that $\Lambda \lesssim M_p$. However, the quartic stabilization term might not always stabilize the potential in both real and imaginary directions of $y_2$. To tackle this problem, we can either include an additional quartic stabilization term in the K\"ahler potential or use the additional term $g(y_1, y_2)$ in the general Branch I~(\ref{genstaro1}) and Branch II~(\ref{genstaro2}) expressions, which does not affect the inflationary potential when $\langle y_2 \rangle = 0$.
Analogously, for the general Branch III and IV expressions we assumed that $\langle y_1 \rangle = 0$, and the stabilization can be achieved by introducing a quartic stabilization term in the K\"ahler potential for the field $y_1$.

Further, it can be readily shown that after fixing the volume modulus field $\langle T \rangle = \frac{k}{2}$ for Branch I or Branch II scalar potentials, the mass of the imaginary component $m_{Im,\phi}^2$ will be positive and independent of the free arbitrary parameter $b$ for the solutions~(\ref{cases1}), while for the complex set of solutions~(\ref{cases2}), $m_{Im,\phi}^2 \geq 0$ will be obtained after setting the arbitrary free parameter $a$ to be purely imaginary, where the value of $|a|$ will determine the curvature in the imaginary direction. Therefore, after fixing the field $T$, the imaginary part of the field $\phi$ will be fixed by the dynamics of the potential to $\langle Im~\phi \rangle = 0$ and this was shown concretely for the Wess-Zumino model in~{\cite{ENO6}}. 
Similarly, for Branch III and IV effective potentials, we fix the matter field $\langle \phi \rangle = 0$. In an identical manner, after fixing the field $\phi$, the imaginary part of the field $T$ will be fixed dynamically to $\langle Im~T \rangle = 0$ and the imaginary mass $m_{Im, T}^2$ will be positive for any free arbitrary parameter $b$ for the four solutions~(\ref{cases1}) and any purely imaginary parameter $a$ for the two solutions~(\ref{cases2}).

One can now observe that the four different branch solutions~(\ref{genstaro1}-\ref{genstaro4}) exhibit similar characteristics. The crucial difference is that for Branch I and II solutions the inflaton is identified with a matter field while for Branch III and IV solutions it is identified with a modulus fields. Therefore, discrete SU(2,1)/SU(2) coset transformations are a powerful tool that can be employed to change the field that will be identified as inflaton, and this will have important consequences on how the inflationary sector couples to matter.  We do not investigate such possibilities here, and the phenomenological aspects of our models will be addressed in the future.

We show next that our general expressions include the SU(2,1)/SU(2)$\times$U(1) no-scale inflationary models considered previously in the literature:
solutions known previously are related through rotations within a branch and/or SU(2,1) transformations.
This classification also allows us to find new, compact forms of superpotential that also yield the Starobinsky potential.

\section{Specific Examples of Starobinsky-Like Models}
\label{examples}

We now show how some specific examples of models yielding a Starobinsky-like effective potential fit within this general classification.

\subsection{Wess-Zumino Superpotential}

It was shown previously~\cite{ENO6} that it is possible to obtain the Starobinsky inflationary potential for a matter field $\phi$ using 
a simple Wess-Zumino superpotential containing only a quadratic and cubic coupling, that can be written as
\begin{equation}
W = M \left[ \frac{\sqrt{k} \phi^2}{2} - \frac{\phi^3}{3 \sqrt{3}} \right],
\end{equation}
where we include a constant $\sqrt{k}$, so that our Wess-Zumino superpotential expression is compatible with the transformation laws~(\ref{symfld1}) and~(\ref{symfld2}). If we switch to the $(y_1, y_2)$ symmetric basis we obtain the following expression \cite{Avatars}:
\begin{equation}\label{wzsym}
W = M \left[\frac{y_1^2}{2} - \frac{y_1^3}{3 \sqrt{3}} + \frac{y_1^2 y_2}{2 \sqrt{3}} \right] \, .
\end{equation}
To recover the Starobinsky inflationary potential, we assume that $y_2$ is fixed so that $\langle y_2 \rangle = 0$, while $W$, $W_1$ and $W_2$ are all non-zero. If one then uses~(\ref{effsym}) and~(\ref{effsym2}), the effective potential becomes:
\begin{equation}
V = \frac{M^2 |y_1|^2 \left|1 - \frac{y_1}{\sqrt{3}} \right|^2}{\left( 1 - \frac{|y_1|^2}{3}\right)^2} = \frac34 M^2 (1 - e^{-\sqrt{2/3}x})^2 \, .
\end{equation}
With the Branch I canonical field redefinition for the symmetric field $y_1$~(\ref{symtrans1}), we obtain the Starobinsky inflationary potential. If we compare the Wess-Zumino superpotential in the symmetric basis~(\ref{wzsym}) to the general Branch I superpotential expression~(\ref{genstaro1}),
we find the following values of the arbitrary coefficients:
\beq
 a, d, e=0; \qquad b = \frac{1}{2}; \qquad c = - \frac{1}{3 \sqrt{3}}; \qquad f = \frac{1}{2 \sqrt{3}} \, ,
 \label{wzgen}
 \eeq 
 which satisfies the general set of coefficient conditions~(\ref{cases1}). 
 
\subsection{Cecotti Superpotential}

The Wess-Zumino model with general coefficients given by Eq. (\ref{wzgen}) is just one particular solution in the general class given by (\ref{cases1}).
We may consider instead the solution with 
\beq
 a, b, c, d=0; \qquad  e = - 1; \qquad f = \frac{1}{\sqrt{3}} \, .
 \label{Cgen}
 \eeq 
We may now perform the SU(2,1) transformation with $\alpha = 0$ and $\beta = -1$ or $y_1 \to -y_2$ and $y_2 \to y_1$ from Branch I to Branch III.
Then, from Eq. (\ref{genstaro3}) we have 
\begin{equation}\label{cec1}
W = M \left[ y_1 y_2  + \frac{y_1 y_2^2}{\sqrt{3}} \right] \, ,
\end{equation}
which when transformed to the $(T, \phi)$ basis gives
\begin{equation}
W = \sqrt{3} M \phi \left(T - \frac{k}{2} \right) \, ,
\end{equation}
which is precisely the Cecotti \cite{Cecotti} form for the superpotential giving rise to the $R+R^2$ theory and Starobinsky inflation when 
$\langle y_1 \rangle = \langle \phi \rangle =0$. 
Indeed, evaluating the scalar potential from (\ref{cec1}) we obtain again the Starobinsky form
\beq
V = \frac{M^2 |y_2|^2 \left|1 + \frac{y_2}{\sqrt{3}} \right|^2}{\left( 1 - \frac{|y_2|^2}{3}\right)^2} = \frac34 M^2 (1 - e^{-\sqrt{2/3} \rho})^2 \, ,
\eeq
using the Branch III field redefinition in (\ref{symtrans3}).

\subsection{Related Superpotentials}
As one can imagine, through a combination of SU(2,1) transformations and choice of solutions from (\ref{cases1}) or (\ref{cases2}),
several other models can be generated.
For example, the transformation with $\alpha = -1$ and $\beta = 0$ or $y_1 \rightarrow -y_1$ and $y_2 \rightarrow -y_2$ takes us from 
Branch I to Branch II or from Branch III to Branch IV.
If we now transform the Wess-Zumino superpotential~(\ref{wzsym}) to Branch II in this way, we obtain
\begin{equation}\label{wzsym2}
W = M \left[ \frac{y_1^2}{2} + \frac{y_1^3}{3 \sqrt{3}} - \frac{y_1^2 y_2}{2 \sqrt{3}} \right] \, .
\end{equation}
Transforming to the $(T, \phi)$ basis, we obtain:
\begin{equation}
W =M \left[
\frac{T \phi ^2}{\sqrt{k}}+\frac{\phi ^3}{3 \sqrt{3}} \right] \, ,
\end{equation}
and the Starobinsky potential is found when $T$ is fixed to $\langle T \rangle = k/2$. 
The Branch III version of the Wess-Zumino model is given by
\begin{equation}\label{wzsym3}
W = M \left[\frac{y_2^2}{2} + \frac{y_2^3}{3 \sqrt{3}} + \frac{y_1 y_2^2}{2 \sqrt{3}} \right] , 
\end{equation}
which, when transformed to the $(T, \phi)$ basis, becomes:
\begin{equation}
W = M \left[
\frac{1}{16k^{3/2}} (k-2 T)^2 \left(2 T+ 2 \sqrt{3} \sqrt{k} \phi +5 k\right) 
\right] \, ,
\end{equation}
as originally found in \cite{Avatars}.
In this case, we must fix $\langle \phi \rangle = 0$ to obtain the Starobinsky potential.
There is a Branch IV analogue also given in \cite{Avatars} as the reversed Wess-Zumino solution with
 superpotential given by:
\begin{equation}\label{wzsym4}
W = M \left[\frac{y_2^2}{2} - \frac{y_2^3}{3 \sqrt{3}} - \frac{y_1 y_2^2}{2 \sqrt{3}} \right] \, .
\end{equation}
Transforming it to the $(T, \phi)$ basis, we find
\begin{equation}
W = M \left[
\frac{1}{16k^{3/2}} (k-2 T)^2 \left(10 T - 2 \sqrt{3} \sqrt{k} \phi +k\right) 
\right] \, .
\end{equation}
Thus, we have identified four different forms of Wess-Zumino superpotential that, by following the corresponding branch field parametrization rules,
yield the Starobinsky inflationary potential. 
All four of these stem from the same solution of (\ref{cases1}) with $b = 1/2$. 
It is also clear from this specific example that some superpotential branches correspond to simpler expressions.
Thus, for a general analysis, it is more convenient to choose the superpotential branch that has the simplest superpotential expression in the $(T, \phi)$ basis.

Similarly, if we take the Cecotti form (\ref{cec1}) and make the same sets of transformations, we obtain solutions in the other branches.
Using  $\alpha = -1$ and $\beta = 0$ or $y_1 \rightarrow -y_1$ and $y_2 \rightarrow -y_2$ we obtain the Branch IV solution
\begin{equation}\label{cec2}
W = M \left[ y_1 y_2  - \frac{y_1 y_2^2}{\sqrt{3}} \right] \, ,
\end{equation}
which becomes \cite{Avatars}
\begin{equation}
W = 
\sqrt{3}M T \phi \left(1 - \frac{2T}{k} \right)
\end{equation}
in the $(T, \phi)$ basis. On the other hand, if we start with the Cecotti superpotential~(\ref{cec1}) and apply the transformations 
$\alpha = 0$, $\beta = -1$, or $y_1 \rightarrow -y_2$ and $y_2 \rightarrow y_1$, 
we obtain the following Branch I superpotential in the symmetric $(y_1, y_2)$ basis:
\begin{equation}\label{cec3}
W = M \left[-y_1 y_2 - \frac{y_1^2 y_2}{\sqrt{3}} \right] \, ,
\end{equation}
and in the $(T, \phi)$ basis we have \cite{Avatars}:
\begin{equation}
W = 
-\frac{1}{4}M \phi  \left(1-\frac{2 T}{k}\right) \left(2 \sqrt{3} T +\sqrt{3} k + 2 \sqrt{k} \phi \right) \, .
\end{equation}
Finally, we give the remaining Cecotti superpotential form, which belongs to Branch II:
\begin{equation}\label{cec4}
W = M \left[-y_1 y_2 + \frac{y_1^2 y_2}{\sqrt{3}} \right] \, .
\end{equation}
If we consider the same superpotential in the $(T, \phi)$ frame, we find:
\begin{equation}
W = 
-\frac{1}{4}M \phi  \left(1-\frac{2 T}{k}\right) \left(2 \sqrt{3} T +\sqrt{3} k - 2 \sqrt{k} \phi \right) \, .
\end{equation}
Therefore, we once again managed to recover four different superpotential forms that with corresponding field fixing and canonical field redefinitions all yield the Starobinsky inflationary potential.

We stress that 
the models highlighted above are relatively simple models based on solutions of Eq. (\ref{cases1}) with the specific values  $b = 0$ or 1/2.
There are of course a continuous family of solutions with $b \in [-1/2, 1/2]$.
Furthermore, the 4 branches we have focused on are obtained from only a discrete subset of the possible SU(2,1) transformations (\ref{syminv}), where we have chosen
$\alpha = 0, -1$ and/or $\beta = 0, -1$. Any pair of values of $\alpha$ and $\beta$ with $|\alpha|^2 + |\beta|^2$ = 1 will yield additional solutions based on Eq. (\ref{cases1}),
so long as the appropriate combination of $y_1$ and $y_2$  ($T$ and $\phi$) is held fixed.

\subsection{Complex Superpotential}
There remains one additional class of solutions of the general Starobinsky superpotential expressions associated with (\ref{cases2}). In this case,  we can build a 
superpotential, where all six coefficients are non-zero and some of the arbitrary coefficients will be complex. 
For example, if we choose the free parameter to be $a = \frac{\sqrt{3}i}{2}$, then the remaining coefficients become: $b=-i;~c = \frac{i}{2 \sqrt{3}};~d=\frac{\sqrt{3}}{2};~e=-2;~f=\frac{\sqrt{3}}{2}$.
With this coefficient choice the general Branch I superpotential~(\ref{genstaro1}) acquires the following form:
\begin{equation}\label{complex1}
W = M 
\left[
\frac{\sqrt{3} i y_1}{2} -i y_1^2 + \frac{i y_1^3}{2 \sqrt{3}} + \frac{\sqrt{3}y_2}{2} -2y_1 y_2+ \frac{\sqrt{3}y_2 y_1^2}{2} 
\right] \, .
\end{equation}
When transforming from the symmetric basis $(y_1, y_2)$ to the $(T, \phi)$ basis, for convenience we will set the coefficient $k=1$ for the transformation laws~(\ref{symfld1}), and obtain:
\begin{equation}\label{complexT}
\begin{split}
W = \frac{1}{48} \left(-18 T \left(4 T^2+2 T-1\right)-12 ((6+4 i) T+(-3+2 i)) \phi ^2 \right. \\
\left.+24 \sqrt{3} T ((4+i) T+i) \phi +8 i \sqrt{3} \phi ^3-(24-6 i) \sqrt{3} \phi +9 \right) \, .
\end{split}
\end{equation}
Although this Branch I superpotential in the $(T, \phi)$~(\ref{complexT}) is somewhat complicated, we can find a simpler form by transforming the Branch I complex superpotential~(\ref{complex1}) to a Branch III superpotential. By using the transformation laws $y_1 \rightarrow -y_2$ and $y_2 \rightarrow y_1$, one obtains:
\begin{equation}
W = M 
\left[
-\frac{\sqrt{3} i y_2}{2} -i y_2^2 - \frac{i y_2^3}{2 \sqrt{3}} + \frac{\sqrt{3}y_1}{2} + 2y_1 y_2 + \frac{\sqrt{3}y_1 y_2^2}{2} 
\right] \, ,
\end{equation}
and then transforming it to the $(T,  \phi)$ basis:
\begin{equation}\label{complex2}
W = \sqrt{3} M\left[\frac{\sqrt{3}i}{4} (2 T-1)
-\phi (T-1) \right] \, .
\end{equation}
Hence, by performing the field transformation, we managed to obtain a compact superpotential form~(\ref{complex2}), which would be a more convenient form if it were chosen as the starting point of the analysis. 

\section{Conclusions}
\label{conx}

We have developed in this paper a general classification of models formulated in the framework of SU(2,1)/SU(2)$\times$U(1)
no-scale supergravity that have a Starobinsky-like effective scalar potential. We have exhibited four different
branches of such models, characterized by different choices of field expectation values and canonical
field redefinitions, as illustrated in Fig.~\ref{fig:branches}. 
These branches are obtained from discrete SU(2,1)/SU(2) transformations where either $\phi$ or $T$ are held fixed.
The branches are in fact related by the continuous set of transformations which require a 
linear combination of $\phi$ and $T$ to be fixed in order to obtain a Starobinsky scalar potential.
Within each branch, there are six classes of
Starobinsky-like models, as shown in Eqns.~(\ref{cases1}) and (\ref{cases2}). The solutions in the
different branches are related via redefinitions of the SU(2,1)/SU(2)$\times$U(1) coset fields, as
illustrated in Fig.~\ref{fig:branches2}. We have also shown how Starobinsky-like models known 
previously~\cite{Cecotti,ENO6,Avatars} are embedded within this general framework, and given some
examples of additional Starobinsky-like models.

Our classification serves as a demonstration that Starobinsky-like inflation is a relatively generic feature
of no-scale supergravity models, unlike simple polynomial models of inflationary potentials, for example.
This may be encouraging for string theorists, in view of the facts that CMB data favour Starobinsky-like models
and that no-scale models emerge as generic low-energy effective field theories derived from string theory.
Until now, the derivation from string theory of a specific superpotential yielding a Starobinsky-like inflationary
model has proved elusive. However, the results of this paper may help by exhibiting the general form of such
superpotentials, thereby extending the target to be aimed at, which is considerably larger than the specific
examples known previously~~\cite{Cecotti,ENO6,Avatars}.

The analysis in this paper may serve as a useful framework for the analysis of present and
future CMB data.  Any deviations from the specific parameter relations in (\ref{cases1}) and (\ref{cases2}) would yield potentially 
observable deviations from the predictions of the Starobinsky model of inflation. Although the present CMB 
data are completely consistent with Starobinsky-like inflation, one should be on the lookout for any possible 
deviations from this paradigm. If observed, they might help identify the context in which Starobinsky-like 
inflation should be embedded. In addition, we want to emphasize that discrete SU(2,1)/SU(2) coset transformations are by no means limited to just the Starobinsky-like inflationary models and the same transformation laws can be successfully applied to any arbitrary models based on a non-compact SU(2,1)/SU(2) coset manifold. Our analysis could serve as a useful guide in this respect. We look forward to the next generation of CMB data following those from the Planck satellite~\cite{planck18}. 

In closing, we recall that there are
two observables in slow-roll inflation, the scalar tilt, including $n_s$ as well as the scalar-to-tensor ratio $r$. Predictions for these quantities are, in general, sensitive
to the number of e-folds of inflation, which depends in turn on the rate of inflaton decay~\cite{EGNODK}. One might expect that this
would be different if the inflaton is identified with a modulus field or a matter field (or some combination of the two). Future measurements
of $n_s$, in particular, could help break the observational degeneracy between different Starobinsky-like models within our
general classification.

\subsection*{Acknowledgements}

\noindent
The work of JE was supported in part by the United Kingdom STFC Grant
ST/P000258/1, and in part by the Estonian Research Council via a
Mobilitas Pluss grant. The work of DVN was supported in part by the DOE
grant DE-FG02-13ER42020 and in part by the Alexander~S.~Onassis Public
Benefit Foundation. The work of KAO was
supported in part by DOE grant DE-SC0011842 at the University of
Minnesota.


\begin{thebibliography}{9}

\bibitem{planck18}
  N.~Aghanim {\it et al.} [Planck Collaboration],
  arXiv:1807.06209 [astro-ph.CO];
  Y.~Akrami {\it et al.} [Planck Collaboration],
  arXiv:1807.06211 [astro-ph.CO].
  
 

  
  \bibitem{reviews}
   K.~A.~Olive,
  Phys.\ Rept.\  {\bf 190} (1990) 307;
A. D. Linde, {\it Particle  
Physics and
Inflationary Cosmology} (Harwood, Chur, Switzerland, 1990); 
  D.~H.~Lyth and A.~Riotto,
{\it Phys.\ Rep.}  {\bf 314} (1999) 1
[arXiv:hep-ph/9807278];
J.~Martin, C.~Ringeval and V.~Vennin,
  Phys.\ Dark Univ.\  {\bf 5-6}, 75-235 (2014)
  [arXiv:1303.3787 [astro-ph.CO]];
  J.~Martin, C.~Ringeval, R.~Trotta and V.~Vennin,
  JCAP {\bf 1403} (2014) 039
  [arXiv:1312.3529 [astro-ph.CO]];
 J.~Martin,
  Astrophys.\ Space Sci.\ Proc.\  {\bf 45}, 41 (2016)
  [arXiv:1502.05733 [astro-ph.CO]].
  
  \bibitem{rlimit}
   P.~A.~R.~Ade {\it et al.} [BICEP2 and Keck Array Collaborations],
  [arXiv:1810.05216 [astro-ph.CO]].
  
  \bibitem{Staro}
A.~A.~Starobinsky,
  Phys.\ Lett.\ B {\bf 91}, 99 (1980).
  
  \bibitem{Nilles:1983ge}
  H.~P.~Nilles,
  Phys.\ Rept.\  {\bf 110} (1984) 1.

\bibitem{Haber:1984rc}
  H.~E.~Haber and G.~L.~Kane,
  Phys.\ Rept.\  {\bf 117} (1985) 75.
  
  \bibitem{natural}
  L.~Maiani,
in Proceedings, Gif-sur-Yvette Summer School On Particle Physics,
  1979, 1-52;
Gerard 't~Hooft and others (eds.),
{\it Recent Developments in Gauge Theories, Proceedings of the Nato Advanced
  Study Institute, Cargese, France, August 26 - September 8, 1979},
Plenum press, New York, USA, 1980, Nato Advanced Study Institutes
  Series: Series B, Physics, 59.;
Edward Witten,
{\em Phys. Lett.} B105, 267, 1981.

  
  \bibitem{EHNOS}
   		H.~Goldberg,
                Phys.\ Rev.\ Lett.\ {\bf 50} (1983) 1419;
                J.~Ellis, J.~Hagelin, D.~Nanopoulos, K.~Olive and M.~Srednicki,
                Nucl.\ Phys.\ B {\bf 238} (1984) 453.

  
  \bibitem{GUTs}
  John~R. Ellis, S.~Kelley and Dimitri~V. Nanopoulos,
{\em Phys. Lett.} B249, 441, 1990;
John~R. Ellis, S.~Kelley and Dimitri~V. Nanopoulos,
{\em Phys. Lett.} B260, 131, 1991;
Ugo Amaldi, Wim de~Boer, and Hermann Furstenau.
\newblock {\em Phys. Lett.}, B260, 447, 1991;
Paul Langacker and Ming-xing Luo,
{\em Phys. Rev.} D44, 817, 1991;
C.~Giunti, C.~W. Kim and U.~W. Lee,
{\em Mod. Phys. Lett.} A6, 1745, 1991.

  
  \bibitem{Hmass}
  H.~E.~Haber and R.~Hempfling,
  Phys.\ Rev.\ Lett.\  {\bf 66} (1991) 1815;
  J.~R.~Ellis, G.~Ridolfi and F.~Zwirner,
  Phys.\ Lett.\ B {\bf 257} (1991) 83;
  Y.~Okada, M.~Yamaguchi and T.~Yanagida,
  Prog.\ Theor.\ Phys.\  {\bf 85} (1991) 1.

  \bibitem{Hcouplings}
  See, e.g., J.~R.~Ellis, S.~Heinemeyer, K.~A.~Olive and G.~Weiglein,
  JHEP {\bf 0301} (2003) 006
  [hep-ph/0211206];
J.~R.~Ellis, S.~Heinemeyer, K.~A.~Olive and G.~Weiglein,
  JHEP {\bf 0301} (2003) 006
  [hep-ph/0211206].
  
  \bibitem{cries}
J.~R.~Ellis, D.~V.~Nanopoulos, K.~A.~Olive and K.~Tamvakis,
  Phys.\ Lett.\ B {\bf 118} (1982) 335.

 \bibitem{no-scale}
E.~Cremmer, S.~Ferrara, C.~Kounnas and D.~V.~Nanopoulos,
  Phys.\ Lett.\ B {\bf 133} (1983) 61.
  
    \bibitem{Ellis:1983sf} 
  J.~R.~Ellis, A.~B.~Lahanas, D.~V.~Nanopoulos and K.~Tamvakis,
  Phys.\ Lett.\  {\bf 134B}, 429 (1984).
 

    \bibitem{LN}
  A.~B.~Lahanas and D.~V.~Nanopoulos,
  Phys.\ Rept.\  {\bf 145} (1987) 1.
  
\bibitem{Witten}
E.~Witten,
  Phys.\ Lett.\  {\bf 155B} (1985) 151.
  doi:10.1016/0370-2693(85)90976-1

   \bibitem{ENO6}
   J.~Ellis, D.~V.~Nanopoulos and K.~A.~Olive,
  Phys.\ Rev.\ Lett.\  {\bf 111} (2013) 111301 
  [arXiv:1305.1247 [hep-th]].

   \bibitem{Cecotti}
S.~Cecotti,
  Phys.\ Lett.\ B {\bf 190} (1987) 86.

\bibitem{Avatars}
   J.~Ellis, D.~V.~Nanopoulos and K.~A.~Olive,
  JCAP {\bf 1310} (2013) 009
 [arXiv:1307.3537 [hep-th]].

\bibitem{EKN1}
 J.~R.~Ellis, C.~Kounnas and D.~V.~Nanopoulos,
  Nucl.\ Phys.\ B {\bf 241}, 406 (1984).
  


\bibitem{EKN2}
 J.~R.~Ellis, C.~Kounnas and D.~V.~Nanopoulos,
  Nucl.\ Phys.\ B {\bf 247}, 373 (1984).
  
    \bibitem{KLno-scale}
R.~Kallosh and A.~Linde,
  JCAP {\bf 1306}, 028 (2013)
  [arXiv:1306.3214 [hep-th]].
  
  
\bibitem{FKR}
 F.~Farakos, A.~Kehagias and A.~Riotto,
  Nucl.\ Phys.\ B {\bf 876}, 187 (2013)
  [arXiv:1307.1137 [hep-th]].


  \bibitem{rs}
D.~Roest and M.~Scalisi,
  Phys.\ Rev.\ D {\bf 92}, 043525 (2015)
  [arXiv:1503.07909 [hep-th]].
  
  \bibitem{GL}
   A.~S.~Goncharov and A.~D.~Linde,
  Class.\ Quant.\ Grav.\  {\bf 1}, L75 (1984).
  
  
\bibitem{EENOS}
J.~R.~Ellis, K.~Enqvist, D.~V.~Nanopoulos, K.~A.~Olive and M.~Srednicki,
  Phys.\ Lett.\  {\bf 152B}, 175 (1985)
  Erratum: [Phys.\ Lett.\  {\bf 156B}, 452 (1985)].


\bibitem{eno9}
  J.~Ellis, D.~V.~Nanopoulos and K.~A.~Olive,
  Phys.\ Rev.\ D {\bf 97}, no. 4, 043530 (2018)
  [arXiv:1711.11051 [hep-th]].

\bibitem{enno}
 J.~Ellis, B.~Nagaraj, D.~V.~Nanopoulos and K.~A.~Olive,
  JHEP {\bf 1811}, 110 (2018)
  [arXiv:1809.10114 [hep-th]].
  
   \bibitem{EKN3}
   J.~R.~Ellis, C.~Kounnas and D.~V.~Nanopoulos,
  Phys.\ Lett.\  {\bf 143B}, 410 (1984).

  
  \bibitem{EGNODK}
  J.~Ellis, M.~A.~G.~Garcia, D.~V.~Nanopoulos and K.~A.~Olive,
  JCAP {\bf 1507} (2015) no.07,  050
  [arXiv:1505.06986 [hep-ph]].
  
 

\end{thebibliography}
\end{document}